# Probing degenerate two-level atomic media by coherent optical heterodyning

**A.M. Akulshin, R.J. McLean, A.I. Sidorov and P. Hannaford**
Centre for Atom Optics and Ultrafast Spectroscopy,
Swinburne University of Technology, Melbourne, Australia

E-mail: aakoulchine@swin.edu.au

**Abstract.** We analyze how light-induced coherent population oscillations and ground-state Zeeman coherence in an atomic medium with degenerate two-level transitions can modify spectra of applied cw resonant radiation at the sub-mW power level. The use of mutually coherent optical fields and heterodyne detection schemes allows spectral resolution at kHz level, well below the laser linewidth. We find that ground-state Zeeman coherence may facilitate nonlinear wave mixing while coherent population oscillations are responsible for phase and amplitude modulation of the applied fields. Conditions for the generation of new optical fields by nonlinear wave mixing in degenerate two-level atomic media are formulated.

1. Introduction

Potential applications in long-distance quantum communication and quantum information processing, as well as fundamental aspects of atom-light interactions, are generating widespread interest in specially prepared coherent atomic media [1,2,3]. To effectively control the properties of such media the spectral and temporal nonlinear processes occurring in them must be understood, since light-induced coherences as well as coherent oscillations of population significantly enhance their nonlinear susceptibility [4,5,6]. While Hanle-type and pump-probe spectral methods have been widely employed for investigations of coherent effects in atoms [1-3,7], we demonstrate that optical heterodyning offers a complementary approach for testing atomic media.

Homodyne detection or self heterodying can be readily demonstrated in alkali atom vapours. Figure 1(a) depicts orthogonally polarized drive and probe components of bichromatic laser radiation propagating through a Rb vapour cell. The frequencies of the two mutually coherent components are offset by $\delta = \nu_D - \nu_P$, which is much smaller than the excited state natural width $\Gamma/2\pi$ of the Rb $D_2$ line. The light intensity is modest, the Rabi frequency $\Omega$ of the drive component being approximately $0.1\Gamma$. Figure 1(b) shows the signal detected by a fast photodiode after the Rb cell. The signal consists of a slow intensity variation due to absorption on the $5S_{1/2}(F=3) \rightarrow 5P_{3/2}$ transitions in $^{85}$Rb and fast oscillations produced by beating between the transmitted optical fields.

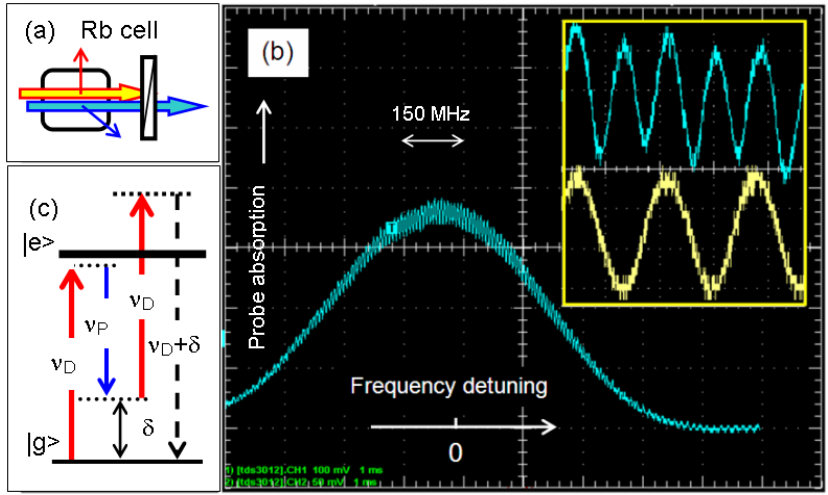

Figure 1. Homodyne detection in a degenerate two-level atomic system. (a) Optical scheme for observation of coherent heterodyning and wave mixing in a Rb cell. (b) Inhomogeneously broadened absorption profile detected by a probe field as a function of frequency detuning from the $5S_{1/2}(F=3) \rightarrow 5P_{3/2}(F'=4)$ transition in $^{85}$Rb. The two curves in the inset are enlargements of the oscillations for resonant (upper) and off-resonant light (lower). (c) Scheme for mixing of two optical fields with frequency offset $\delta$ that generates a new optical field at $\nu_N=\nu_D+\delta$.

For non-resonant light, when the polarizer allows both orthogonal polarizations to pass, the oscillations are closely fitted by a sine wave of frequency $\delta$. Oscillations are much smaller if the drive component is blocked by the polarizer. For the resonant case, however, when the laser radiation is scanned across an absorption component of the Rb $D_2$ line, oscillations at $2\delta$ appear. The oscillations are stronger at the top of the Doppler profile and are evidence of new optical fields generated in the Rb cell as a result of four-wave mixing (FWM). Mutual coherence of the applied optical fields is crucial for getting such monochromatic beat signals, otherwise the spectrum is determined by a convolution of linewidths of two independent optical fields. As shown schematically in figure 1(c) the absorption of two photons from the drive field with frequency $\nu_D$ and stimulated emission at $\nu_P$ gives rise to a new optical wave with electric field amplitude $E_N \sim \chi^{(3)} E_D^2 E_P$, where $E_D$ are $E_P$ are the amplitudes of the bichromatic components and $\chi^{(3)}$ is the nonlinear susceptibility of the atomic medium. The new wave at $\nu_N = 2\nu_D-\nu_P = \nu_D+\delta$ when mixed on the photodiode with the component at $\nu_P = \nu_D-\delta$ produces a beat signal at $2\delta$, suggesting that the new field is also mutually coherent with the applied radiation. This simple experiment illustrates that FWM in an atomic medium can be very efficient. Indeed, it may often occur when it is not intended, changing spectral and amplitude properties of the applied resonant radiation in pump-probe schemes.

Four-wave mixing is one of the most studied processes in nonlinear optics [8]. In a strongly driven two-level atomic system FWM can be enhanced by the ac-Stark effect as was shown in [9,10], while in an alkali atomic vapour it can be further enhanced in the vicinity of a Raman transition. Previous studies have demonstrated that light-induced coherence between ground-state sublevels, which is responsible the effects of coherent population trapping (CPT) [11] and electromagnetically induced transparency (EIT) [12], may facilitate the generation of new optical components detuned from the applied radiation by the ground-state splitting [13,14,15]. It was shown that light propagation in EIT atomic media with a double-$\Lambda$ level configuration is affected by resonant FWM [16].

Light-induced Zeeman coherence within degenerate ground and excited levels of an atomic transition leads to different consequences for the dissipative properties of the atomic medium depending on the degeneracy of the levels involved. The interaction of monochromatic light and atoms with transitions of the type $F_g \rightarrow F_e = F_g -1$ and $F_g \rightarrow F_e = F_g$, where $F_g$ and $F_e$ are the total angular momentum of the ground and excited levels, can produce a coherent superposition of ground-level Zeeman states that is decoupled from the light field [17], causing the absorption of resonant light to be reduced due to CPT. The situation is different for an $F_g \rightarrow F_e = F_g+1$ cycling transition. In this case the medium under the action of resonant radiation may exhibit enhanced absorption above the linear absorption level, or electromagnetically induced absorption (EIA) [18]. EIA is explained in terms of transfer of ground-state coherence between the ground and excited levels [19,20]. Despite this remarkable difference in dissipative properties, enhanced and highly selective Kerr nonlinearity occurs in the sub-MHz vicinity of a Raman transition for both EIT- and EIA-type degenerate two-level atomic media [21].

Considerable attention has been paid to backward FWM in degenerate or nearly degenerate atomic media [22], while forward FWM in two-level EIT media has been used for the generation of correlated photon pairs [23] and intensity-squeezed light fields [24]. Bragg diffraction by an electromagnetically induced grating has been studied in an ensemble of cold degenerate two-level atoms [25,26]. Numerical calculations of FWM spectra enhanced by Zeeman coherence in a degenerate two-level system were presented by Lezama [27] using a semiclassical treatment of the atom-light interaction.

In this paper we present results of an optical heterodyning study of wave mixing enhanced by long-lived Zeeman coherences and efficient phase modulation produced by coherent population oscillations (CPO) in Rb vapour. Our aim is to illustrate the rich potential this approach offers for understanding the processes involved in generating enhanced atomic Kerr nonlinearities. In particular, we show that this technique also offers a method for distinguishing the contributions made by long-lived Zeeman coherence and CPO to the laser-induced nonlinearity of the atomic medium.

2. **Method**

New optical fields generated in an atomic medium with a coherent superposition of different ground-state hyperfine levels have been easily resolved using a Fabry-Perot (FP) interferometer because of their large frequency offset, of the order of the ground-state splitting, from the applied laser radiation [14]. In the case of FWM in degenerate two-level atomic media, new optical fields have much small frequency separations from the applied laser radiation, typically less than the natural linewidth of the optical transition. To resolve the new fields using conventional spectroscopic tools, such as a FP cavity, exceptionally high temporal coherence of the laser fields is required in addition to better than 100 kHz spectral resolution of the cavity. However, the application of heterodyne methods in combination with mutually coherent optical fields eases this demand. The ultimate resolution of these methods is limited by the mutual coherence, or frequency noise correlation of the bichromatic components, rather than by absolute linewidths of the applied resonant light.

We use homodyne (or self-heterodyne) and heterodyne methods to investigate the wave mixing process and analyze changes in the frequency spectra of optical fields transmitted through the atomic medium with very high spectral resolution. Both the transmitted laser radiation and the new optical fields resulting from wave mixing contribute to the detector output. In both methods, the photodiode output is processed by a radio frequency spectrum analyzer.

In the homodyne method, the signals arise from beating between the applied laser radiation and the newly generated fields, while in the more versatile heterodyne detection method a reference (local oscillator) wave is combined with the radiation leaving the cell. The heterodyne method is also more sensitive, because the amplitude of the beat signal is proportional to the amplitudes of the reference field $E_R$, which can be made arbitrarily strong, resulting in an enhancement of approximately $(I_R/I_1)^{1/2}$ in the beatnote. Also it is not necessary to have either a polarizer or an angle between the drive and probe components to separate the drive and probe components.

Typically the frequency offset of the reference wave from the bichromatic radiation entering the cell is much larger than the width of the coherence resonances. However, the reference wave should still be coherent with the applied laser radiation to preserve the sub-linewidth spectral resolution, and so in this work is derived from the same laser.

Thus, both homodyne and heterodyne methods allow very sensitive detection of variations in the optical field spectra produced by resonant atom-light interaction.

3. **Experimental set-up**

A simplified scheme of the experimental setup is shown in figure 2. The glass cell is 5 centimeters long and contains a natural mixture of Rb isotopes and no buffer gas. All experiments presented in this paper are performed on the $^{85}$Rb isotope. A high-permeability shield reduces ambient magnetic fields to less than 5 mG when required. The temperature of the cell is set within the range 30-50$^\circ$C.

A single-mode extended-cavity diode laser tuned to the Rb-D$_2$ line is used as the resonant light source. The linewidth of the laser is about 1 MHz. Two acousto-optic modulators (AOM) are used to produce two mutually coherent fields with optical frequencies $v_1$ and $v_2$ having a small tunable offset ($\delta = v_2 - v_1 << \Gamma/2\pi \approx 6$ MHz), by successively shifting the optical laser frequency $v_L$ up by 80 MHz then down by ($\delta$+80) MHz. The mutual coherence of the two beams is determined by the frequency noise added by the AOMs. The two beams are carefully combined into a bichromatic beam and passed through the cell. The divergences of the components of the bichromatic beam are closely matched to avoid geometrical broadening of the subnatural absorption resonances. The polarizations of the beams are controlled via the polarizers and wave plates. Typically the power of both the bichromatic components is less than 0.4 *mW*, while the power of the reference beam is approximately 0.7 *mW*. The bichromatic beam cross section is in the range 1-6 mm$^2$.

Wave mixing signals detected by the fast photodiode (PD) are analyzed using a radio frequency spectrum analyzer when a fixed frequency offset $\delta$ is applied, while for the probe-pump method, where the offset is scanned around a $\delta=0$ offset, signals are observed and recorded using a digital oscilloscope. In the heterodyne method the unshifted beam after the second AOM at ($v$+80) MHz mixed with radiation leaving the cell acts as a local oscillator. The mutual coherence of the beams is determined by the frequency noise added by the AOMs. The ultimate FWHM of the beat signal at 80 MHz is 700 Hz, whereas the beat signal at (80+$\delta$) MHz is wider, approximately 10 kHz, due to additional noise introduced by the AOM-2 driver.

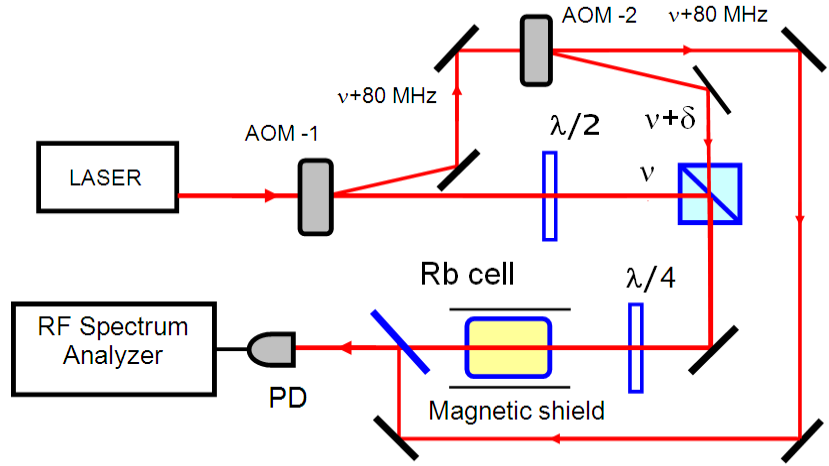

Figure 2: Experimental arrangement. The beam at ($v$ + 80) MHz is the reference beam for the heterodyne arrangement, not used for homodyne experiments.

## 4. Results

First we consider the case when the components of the applied bichromatic radiation have orthogonal linear polarizations. This configuration allows the generation of long–lived Zeeman coherences that enhance the wave mixing process. Also, for orthogonal linear polarizations sub-natural coherent absorption resonances have the highest contrast resulting in the largest value of frequency-selective Kerr nonlinearity [20].

We compare four-wave mixing in EIA- and EIT-type media. The coherent superposition responsible for EIA is prepared by bichromatic radiation tuned to the $5S_{1/2}(F=3) \rightarrow 5P_{3/2}(F'=4)$ cycling transition in $^{85}$Rb while EIT or dark superposition is generated on the $5S_{1/2}(F=2) \rightarrow 5P_{3/2}(F'=1, 2)$ transitions.

Typical EIA- and EIT-type behaviour of Rb vapour in the vicinity of the Raman transition is demonstrated in Fig. 3. In both cases, as was pointed out in [28], the atom-light interaction occurs within three different velocity groups independently; however, the contributions from the cycling transitions dominate because of optical pumping.

Figures 3a, 3c and 3e show normalized transmission of the $v_1$ component as a function of frequency detuning $\delta$ between the components of the bichromatic radiation. The $v_2$ component with orthogonal polarization is blocked after the cell by the polarizer. The typical FWHM of the transmission resonances is smaller than the natural linewidth of the optical transitions within the Rb D$_2$ absorption line. In the vicinity of both EIT and EIA resonances the intensity dependent refractive index results in enhanced Kerr nonlinearity of the Rb vapour.

Figures 3b, 3d and 3f show homodyne detection spectra for a fixed detuning $\delta$, for EIA and EIT media. The individual peaks are easily resolved because of the high mutual coherence of the components of the bichromatic field and their widths are not affected by power broadening of the optical transitions.

Although one might not expect orthogonally polarized fields to beat together to generate a signal at $\delta = v_2 - v_1$, we do observe small beat signals because of imperfect orthogonality of the polarizations resulting from dichroism and birefringence occurring in the optical elements of the apparatus and in the atomic medium. For opposite circular polarizations (Fig. 3f) the peak at $\delta$ is higher than for orthogonal linear polarizations (Fig. 3d) which we attribute to the more severe degradation of the circular polarizations at the mirrors. The signal at $\delta$ is at least two orders of magnitude stronger if the components of the bichromatic beam have parallel polarizations as shown in Fig. 5.

Beating of the components of the bichromatic beam generates a signal at $\delta$ even for orthogonal polarizations and in the absence of atomic media. The signal at $2\delta$, on the other hand, does require atom-light interaction. It can arise in two ways: as a result of beating of the new optical field at $2v_1-v_2 = v_1-\delta$ with the component at $v_2=v_1+\delta$ or as a result of mixing of the field at $2v_2-v_1 = v_1+2\delta$ with the component at $v_1$. These are likely to be comparable in strength, particularly if the applied components at $v_1$ and $v_2$ have similar intensities. We also have to emphasize that despite the observed width of the $2\delta$ peaks of the order of 10 kHz, the absolute linewidth of the new optical fields is limited by the laser linewidth, which is approximately 1 MHz in our case.

The spectra of Figs. 3b and 3d demonstrate that the four-wave mixing is more efficient in the EIA medium than in the EIT medium under otherwise similar experimental conditions, with the peak at $2\delta$ much higher for the EIA case. This observation is consistent with the values of the Kerr coefficients $n_2$ measured directly in [20], where higher values were found for EIA media. It also indicates that a long ground-state relaxation time rather than ultraslow light propagation itself is important for efficient wave mixing.

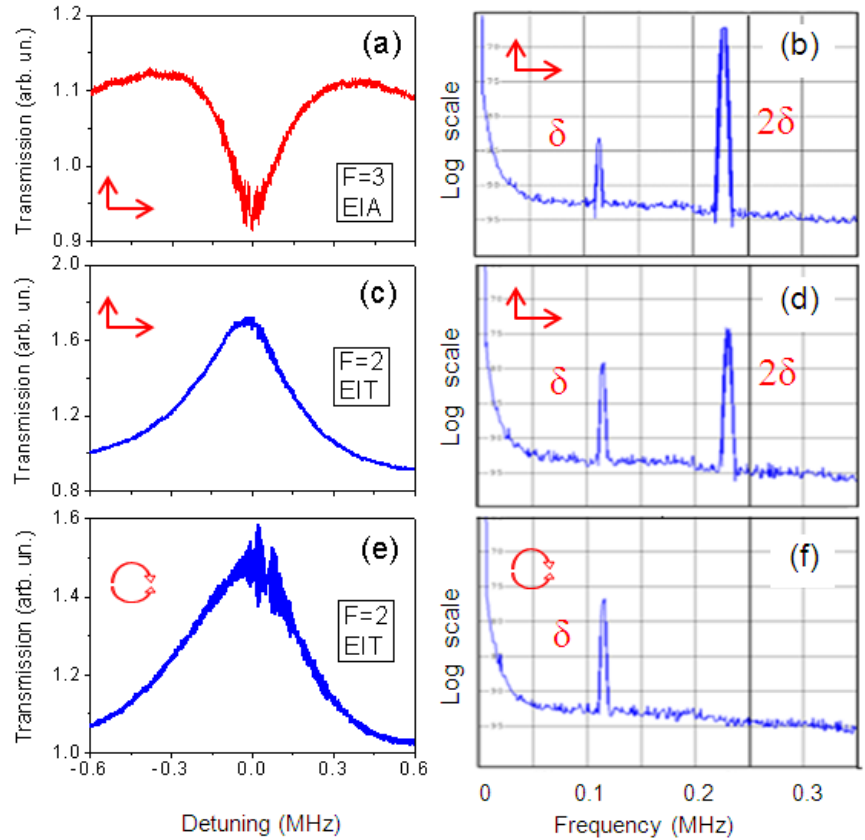

Figure 3. (Left) Coherent transmission resonances as a function of frequency detuning between the components of the bichromatic radiation.
(Right) Homodyne spectra taken under EIT or EIA conditions in Rb vapour. Bichromatic components have fixed frequency offset $\delta \approx 112$ kHz. Vertical and horizontal scales are 5 dBm/div and 50 kHz/div, respectively.
(a, b) and (c, d): Components of bichromatic radiation with orthogonal linear polarizations tuned to the $5S_{1/2}(F=3) \rightarrow 5P_{3/2}$ and $5S_{1/2}(F=2) \rightarrow 5P_{3/2}$ transitions, respectively. (e, f): Components of bichromatic radiation with opposite circular polarizations tuned to the $5S_{1/2}(F=2) \rightarrow 5P_{3/2}$ transition.

The requirements for the observation of FWM and narrow absorption resonances due to Zeeman coherence are different, as was mentioned in [25]. Despite the similarity in the transmission spectra in Figs. 3c and 3e, obtained in EIT-type media with orthogonal linear and opposite circular polarizations, the homodyne spectra differ radically: there is no peak at *2δ* for the case of the opposite circular polarizations as they do not allow four wave mixing to occur. Indeed, the combination of $\sigma^+$ and $\sigma^-$ transitions would require the four-wave mixing component at *2δ* to satisfy a change *Δm=±3* in the magnetic quantum number, as illustrated in figure 4a, which is not allowed by angular momentum conservation. By contrast, linear orthogonal polarizations of the bichromatic components allow $\pi-\sigma^{\pm}-\pi$, $\sigma^+-\pi-\sigma^-$ or $\sigma^--\pi-\sigma^+$ combinations of optical transitions that do give four wave mixing components at *2δ*, as shown in figure 4b,c. Only these combinations of $\pi$ and $\sigma$ transitions can produce coherence between different ground-state magnetic sublevels, satisfy angular momentum conservation, and generate fields with different optical frequencies. Thus, for new field generation by resonant wave-mixing in atomic media both enhanced Kerr nonlinearity and angular momentum conservation are required.

In figure 5 we compare beat spectra for parallel and orthogonal linear polarizations of the bichromatic components tuned to the $^{85}$Rb $5S_{1/2}(F=3) \rightarrow 5P_{3/2}(F'=4)$ cycling. In the case of parallel polarizations, in addition to the expected enhancement of the peak at $\delta$, a peak at $3\delta$ appears.

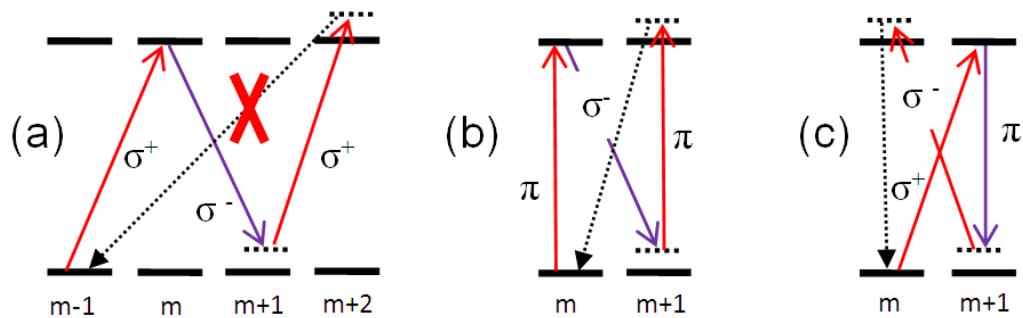

Figure 4. Schemes for four-wave mixing in degenerate two-level media that (a) do not lead to in and (b) lead to generation of fields frequency detuned by $2\delta$ from one of the bichromatic components for (a) $\sigma^+$–$\sigma^-$, (b, c) $\pi$–$\sigma^\pm$ transitions at $v_1$ and $v_2$, respectively. Dotted lines represent new fields.

The origin of the $3\delta$ peak is an important aspect of this work. While in principle the new waves at $2v_1-v_2 = v_1-\delta$ and $2v_2-v_1 = v_1+2\delta$, which are comparable in strength, could beat together to produce a signal at $3\delta$, this signal is too weak to be detected for the case of orthogonal polarizations as shown in Fig. 5. Nonlinearity higher than $\chi^{(3)}$ could also be responsible for this peak; however we attribute it to phase modulation of the bichromatic radiation rather than higher order wave mixing, as described below.

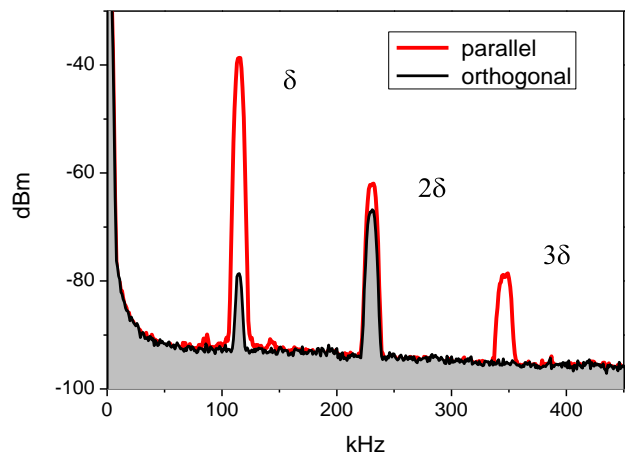

Figure 5. Homodyne spectra taken in EIA atomic media for linear parallel and orthogonal polarizations of the component of the bichromatic radiation.

While sub-natural absorption resonances observed in atomic media are usually associated with ground-state coherence, the excitation of an open, purely two-level atomic system by co-propagating optical waves with a tunable frequency detuning $\delta = v_2-v_1$ can also result in a sub-natural resonance [29]. The periodic modulation of the ground-state population at $\delta$ and multi-photon Raman scattering are responsible for this effect, known as coherent population oscillation (CPO). Including CPO in the description of coherent nonlinear processes in a strongly driven resonant system has made possible a more complete evaluation of the nonlinear susceptibility [30]. CPO has also been exploited to produce slow light in solid-state media [31]. However, the effect has been seldom discussed in connection with experimentally obtained narrow absorption resonances in atomic media because in real atoms with a complex energy level structure the effects of two-level CPO are generally masked by those of light-induced ground-level coherence. It has nevertheless been recently demonstrated that with the pump-probe method under appropriate experimental conditions it is possible to distinguish sub-natural resonances produced by long-lived Zeeman coherence and CPO [32].

Harmonic oscillations of population at $v_2-v_1$ driven by the bichromatic field result in oscillation of both the real and imaginary parts of the atomic susceptibility $\chi = \chi' + i\chi''$, which are responsible for the refractive index $n(v) = 2\pi\chi'(v)$ and the absorption coefficient $\kappa(v)=8\pi^2 v\chi''(v)$ of the medium. Propagation of the initially monochromatic optical component $E_1=A_1\cos(2\pi v_1 t+ k_1 z)$ with field amplitude $A_1$ through the medium with oscillating refractive index $n(t)= n_0(v)+\Delta n \sin(2\pi\delta t)$ and absorption $\kappa(t)= \kappa_0(v)+\Delta\kappa \sin(2\pi\delta t)$, where $n_0(v)$ and $\kappa_0(v)$ are the time-independent components in the vicinity of the Raman transition, produces phase and amplitude modulation of the resonant light and modifies the spectrum of the optical field.

Phase modulation of a sine wave yields sidebands spaced by the modulation frequency $\delta$ and with amplitudes given by the n[th]-order Bessel function $J_n(m)$, so that the spectrum of the phase modulated $E_1$ component can be expressed as

$$E_1 = A_1 J_0 \cos 2\pi v_1 t - J_1(m) \times \cos 2\pi(v_1 - \delta)t - \cos 2\pi(v_1 + \delta)t + J_2(m) \times \cos 2\pi(v_1 - 2\delta)t - \cos 2\pi(v_1 + 2\delta)t - \cdots,$$

where $m=2\pi \Delta n L/\lambda_1$ is the modulation index and $L$ is the length of the Rb cell. The vector sums of odd-order sideband pairs are always in quadrature with the carrier component, while the vector sums of the even-order sideband pairs are always in phase with the carrier.

The amplitude modulation produces two sidebands, which are always in phase with the carrier
$$E_1 = A_1 \cos 2\pi v_1 t + \frac{M}{2}[\cos 2\pi(v_1 + \delta)t + \cos 2\pi(v_1 - \delta)t],$$
where $M$ is the amplitude modulation index.

If the modulation indices $m$ and $M$ are small ($\Delta nL/\lambda<<1$ and $M<<1$) so that only first order sidebands have significant amplitudes, then independent phase and amplitude modulations produce indistinguishable, symmetric spectra because the beat signal does not contain phase information. On the other hand sideband asymmetry indicates that phase and amplitude modulations exist simultaneously at the same frequency.

The $3\delta$ peak in Figure 5 is the result of mixing of CPO-induced sidebands of the bichromatic components. As both the $E_1$ and $E_2$ components of the bichromatic beam are phase and amplitude modulated by the atomic medium, the $3\delta$ peak could result either from mixing of the first low frequency sideband of the $E_1$ component at $v_1-\delta$ and the first high frequency sideband of the $E_2$ component at $v_2+\delta = v_1+2\delta$, or from mixing of the $J_2(m)$ sideband of the $E_2$ component at $v_2+2\delta$ with the $E_1$ component at $v_1=v_2-\delta$. If higher order sidebands are considered, many more possibilities arise for generating the $3\delta$ peak.

CPO is essentially a two-level effect, and for all the combinations of polarizations of the bichromatic radiation the atom-light interaction investigated in this work is reduced to either $\pi$-$\pi$, $\sigma^+$-$\sigma^+$ or $\sigma^-$-$\sigma^-$ transitions for both components. These transitions are unable to generate Zeeman coherence between different ground-state magnetic sublevels.

External magnetic fields influence the shape of the observed spectra. While other descriptions are valid, we can think of this as the external magnetic field imposing a quantization axis and hence determining the relative strength of optical $\sigma^\pm$ or $\pi$ transitions between the ground and excited magnetic sublevels. Figure 6 illustrates this effect for the case of orthogonal linear polarizations of the bichromatic components. The beat signal depends strongly on whether or not the Rb cell is shielded from the ambient magnetic field. With no magnetic shielding around the cell, we expect the magnetic field of the order of 1 gauss to be randomly oriented with respect to the propagation direction of the light, and all $\sigma^+$, $\sigma^-$ and $\pi$ transitions become possible for the both components simultaneously, as was shown, for example, in [12,19]. The $\sigma^+$-$\sigma^+$, $\sigma^-$-$\sigma^-$ and $\pi$-$\pi$ transitions driven by the optical components at $v_1$ and $v_2$ result in CPO generation and the appearance of sidebands due to phase and amplitude modulation and the peak at $3\delta$ is the result of the spectral modification of the bichromatic radiation due to coherent oscillations. In the case of the shielded cell, if the quantization axis is taken to be in the direction of the wave vectors of the applied bichromatic beam, then although both components can induce $\sigma^{\pm}$ transitions, it is not possible for them to generate CPO, because of the $\pi/2$ phase shift between the components. This leads to inefficient phase modulation and to undetectable amplitudes of the frequency sidebands. The magnetic field dependence of the peak at $3\delta$ is convincing evidence of the CPO contribution.

The experimental arrangement did not allow a detailed exploration of the magnetic field dependence of the $3\delta$ peak amplitude. The observation of the $3\delta$ peak in the unshielded magnetyic field of approximately 1 gauss is consistent with it originating from a CPO effect, since ground state atomic coherence effects should occur at magnetic fields causing Zeeman shifts less than the corresponding decoherence rate (of the order of 500 kHz) while CPO effects should persist for Zeeman shifts up to the homogeneous linewidth (several MHz, corresponding to several gauss). This is an interesting area for further investigation.

Finally we note that the $2\delta$ and $3\delta$ peaks shown in Figure 6 are wider than the peak at $\delta$ because of high-index frequency modulation of the $E_2$ component at 50 Hz introduced by the AOM driver.

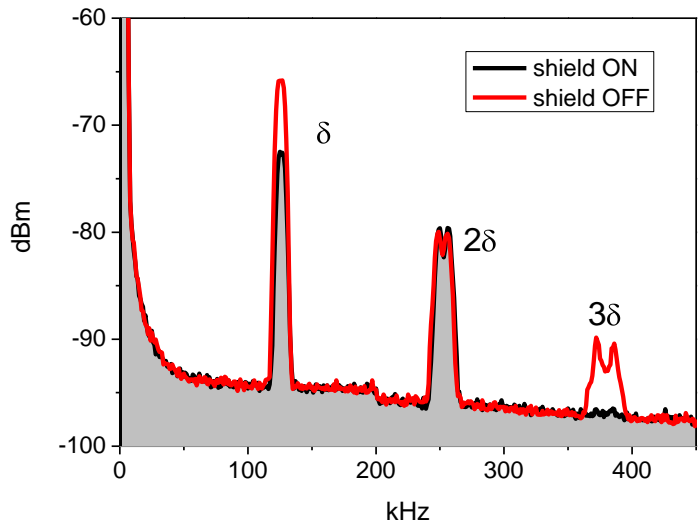

Figure 6. RF spectra for transmitted light tuned to the $5S_{1/2}(F=3) \rightarrow 5P_{3/2}(F'=4)$ transition, EIA-type atomic medium, for orthogonal linear polarizations of the component of the bichromatic radiation for shielded and unshielded Rb cell.

The interpretation of homodyne spectra is not always entirely straightforward. For example, the signal at $\delta$, as well as containing a contribution from the beating of the two bichromatic components, also contains four-wave mixing or CPO components. Both the new wave generated at $2v_2-v_1 = v_2+\delta$ beating with the field at $v_2$ and the new wave generated at $2v_1-v_2 = v_1-\delta$ beating with the field at $v_1$ can generate signals at $\delta$. The fact that individual peaks in the spectra can have several contributions complicates the interpretation of, for example, power dependences of the wave mixing process.

The heterodyne detection method produces spectra that can be more readily interpreted. In this case an additional reference beam shifted by approximately 80 MHz mixes with the radiation leaving the cell to produce beat signals separated by $\delta$. These signals are generated even with off-resonant light, requiring no atom-light interaction. For resonant light the number of peaks and their intensities depend on the atomic transitions, atomic density, and the intensity and polarization of the resonant optical fields. At low light intensity each additional peak in the spectrum corresponds to a new wave generated in the atomic cell. This makes heterodyne spectra easier to interpret than homodyne spectra.

Figure 7 shows a spectrum obtained with heterodyne detection for light propagating through an EIA medium. In addition to the beat signals between the reference beam and the applied radiation, strong additional peaks at $\pm\delta$ are observed for the orthogonally polarized components of the bichromatic radiation.

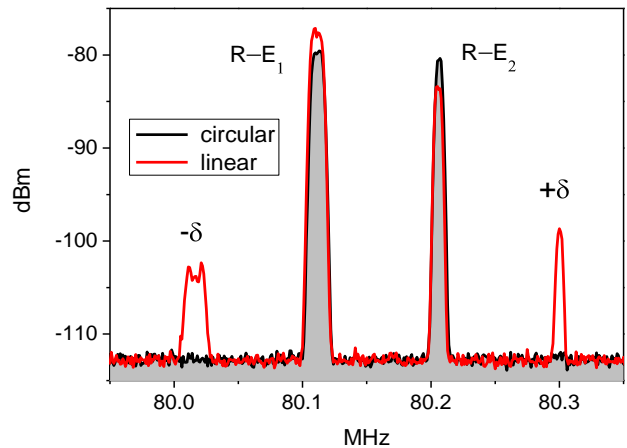

Figure 7. Heterodyne four-wave mixing spectra for linear orthogonal and circular opposite polarizations in an EIA medium shielded from stray magnetic fields. The reference beam is linearly polarized at 45 degrees with respect to the polarizations of the bichromatic radiation. The peaks labeled R-$E_1$ and R-$E_2$ are the result of beating between the reference beam and the components of the bichromatic beam, while the peaks at $\pm\delta$ are due to mixing between the new waves and the reference beam.

The widths of the peaks indicate that the spectral resolution of the heterodyne detection scheme is not limited by the laser linewidth, and that both new waves have a high degree of mutual coherence with the reference beam (and hence with the bichromatic radiation). However, the widths of the peaks shown in Fig. 7 vary. This is due to an additional frequency noise introduced by the AOM-2 driver. The R-$E_1$ peak is broader than the R-$E_2$ peak because the frequency noise on the $E_1$ component of the bichromatic radiation is higher than on the $E_2$ component. The difference between the peaks at $-\delta$ and $+\delta$ is even more pronounced. This is because of the fact that the broader peak at $-\delta$ results from a FWM process involving two photons from the noisier component at $\nu_1$ and one from the more coherent field at $\nu_2$. The deliberate introduction of noise to the AOM driver allows the origin of each heterodyne peak to be understood.

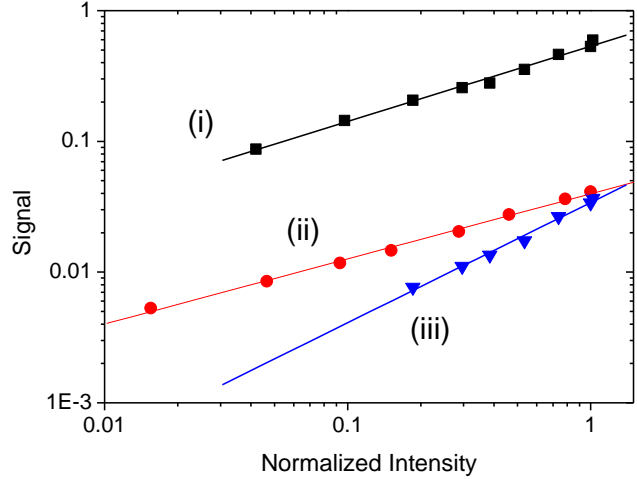

Figure 8. Intensity dependences of the heterodyne peaks shown in Fig. 7. Curve (i) demonstrates how the amplitude of the R-$E_1$ peak depends on normalized intensity of the $\nu_1$ component. Curves (ii) and (iii) show the intensity dependence of the -$\delta$ peak on the normalized intensity of the $\nu_2$ and $\nu_1$ components, respectively. Slopes of the fitted lines are 0.57±0.02; 0.59±0.03 and 0.92±0.01, respectively.

The intensity dependence of the beat signals provides another way of distinguishing the mechanisms responsible for their origin. The power in each spectral component depends on a number of parameters, including the power of the applied radiation. Several such dependences are plotted in Fig. 8 on a log-log scale for the peaks shown in figure 7. Two of the dependences have a slope of approximately one half, while for the third one the slope is close to unity. The -$\delta$ peak arises from beating of the new field at $\nu_1$-$\delta$, with amplitude $E_N \sim \chi^{(3)} I_1 (I_2)^{1/2}$, and the reference, so that its amplitude is proportional to the product $E_N E_R \sim \chi^{(3)} I_1 (I_R I_2)^{1/2}$, implying a linear dependence on the intensity $I_1$ of the $\nu_1$ component, and a square root dependence on the intensity $I_2$, which is consistent with the observed dependences.

## 5. Conclusion

We have shown that two mutually coherent, co-propagating, sub-mW optical waves with a frequency detuning that is much less than the natural linewidth of the optical transitions of the Rb $D_2$ line may produce mutually coherent optical waves in Rb vapours.

The spectra of the radiation transmitted through the atomic medium have been investigated using both homodyne and heterodyne methods. The use of mutually coherent optical fields allows spectral details of the applied and new optical fields to be resolved to a level well below the laser linewidth. The heterodyne scheme, in which a third beam acts as a local oscillator, produces signals that are more readily interpreted. The spectra have different appearances for media possessing EIT and EIA, and for different polarizations of the applied resonant bichromatic laser radiation.

We find that ground-state Zeeman coherence leads to nonlinear wave mixing while coherent population oscillations are responsible for phase and amplitude modulation of the applied fields. Coherent population oscillations, which have not been widely considered in this context before, play a crucial role in determining the observed spectra. The $\sigma^+$-$\sigma^+$, $\sigma^-$-$\sigma^-$ and $\pi$-$\pi$ transitions, driven by the bichromatic field, result in CPO generation and the appearance of sidebands due to phase and amplitude modulation. The $\pi-\sigma^{\pm}-\pi$, $\sigma^+-\pi-\sigma^-$ or $\sigma^--\pi-\sigma^+$ combinations can produce coherence between different ground-state magnetic sublevels, satisfy angular momentum conservation, and generate fields with different optical frequencies. We demonstrate that angular momentum conservation plays a crucial role in FWM in addition to enhanced Kerr nonlinearity.

As coherent heterodyning can be easily implemented with other alkali atoms, this approach represents a novel and complementary method for testing atomic media, combining very high spectral resolution, enhanced sensitivity compared to the pump-probe method (by a factor of approximately $(I_R/I_1)^{1/2}$) and strong spatial selectivity because of the single-mode nature of heterodyne detection.

The results and ideas suggested could be crucial for the modelling of comb-like spectra generated in vapours of alkali atoms and qualitative comparison with experiments. Our study could extend the physical basis of the efficient control of wave mixing in coherently driven atomic media and its application to non-classical light generation.

**Acknowledgements**

We thank Will Brown for his assistance with the experiments.